\RequirePackage{fix-cm}
\documentclass{svjour3}                     
\smartqed  

\usepackage[utf8]{inputenc}
\usepackage[english]{babel}
\usepackage{amsmath}
\usepackage{amsfonts}
\usepackage{amssymb}
\usepackage{graphicx}
\usepackage{tikz}
\usepackage{pgfplots}
\usepackage{xcolor}
\usepackage{textcomp}
\usepackage{url}
\usepgfplotslibrary{units}
\usetikzlibrary{pgfplots.statistics}
\definecolor{d2}{rgb}{1.0,0.0,0.0}
\definecolor{d4}{rgb}{0.7,0.0,0.0}
\definecolor{d6}{rgb}{0.4,0.0,0.0}
\definecolor{d8}{rgb}{0.1,0.0,0.0}

\begin{document}

\title{Learning Soft Tissue Behavior of Organs for Surgical Navigation with Convolutional Neural Networks}

\titlerunning{Learning Soft Tissue Behavior}        

\author{Micha Pfeiffer \and
		Carina Riediger \and
		Jürgen Weitz \and
		Stefanie Speidel}


\institute {Micha Pfeiffer (micha.pfeiffer@nct-dresden.de) \and Stefanie Speidel \at National Center for Tumor Diseases (NCT), Partner Site Dresden, Germany
\and Carina Riediger \and Jürgen Weitz \at Department for Visceral, Thoracic and Vascular Surgery, University Hospital, Technical University Dresden}

\date{Received: date / Accepted: date}

\maketitle

\begin{abstract}
Purpose:
In surgical navigation, pre-operative organ models are presented to surgeons during the intervention to help them in efficiently finding their target.
In the case of soft tissue, these models need to be deformed and adapted to the current situation by using intra-operative sensor data. A promising method to realize this are real-time capable biomechanical models.

Methods:
We train a fully convolutional neural network to estimate a displacement field of all points inside an organ when given only the displacement of a part of the organ's surface.
The network trains on entirely synthetic data of random organ-like meshes, which allows us to generate much more data than is otherwise available.
The input and output data is discretized into a regular grid, allowing us to fully utilize the capabilities of convolutional operators and to train and infer in a highly parallelized manner.

Results:
The system is evaluated on in-silico liver models, phantom liver data and human in-vivo breathing data. We test the performance with varying material parameters, organ shapes and amount of visible surface.
Even though the network is only trained on synthetic data, it adapts well to the various cases and gives a good estimation of the internal organ displacement.
The inference runs at over 50 frames per second.

Conclusions:
We present a novel method for training a data-driven, real-time capable deformation model. The accuracy is comparable to other registration methods, it adapts very well to previously unseen organs and does not need to be re-trained for every patient.
The high inferring speed makes this method useful for many applications such as surgical navigation and real-time simulation.


\end{abstract}


\begin{figure}
\begin{tikzpicture}
	\node[anchor=south west,inner sep=0] at (0,0) {
   		\includegraphics[width=\textwidth]{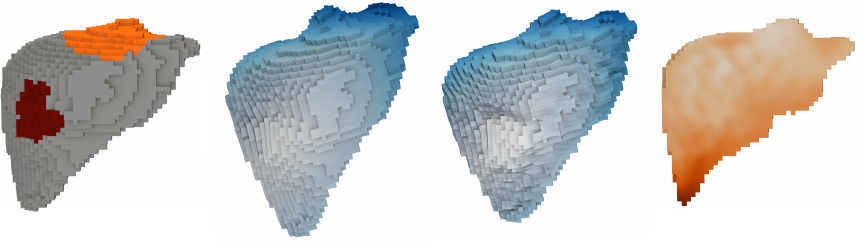}
    };
    \node[anchor=south west,inner sep=0] at (4.8,-0.22) {
   		\includegraphics[width=2cm]{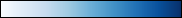}
    };
	\node[text width=3cm] at (5.55,-0.12) {$0$ cm};
	\node[text width=3cm] at (8.45,-0.12) {$8$ cm};
	
	\node[anchor=south west,inner sep=0] at (9.2,-0.22) {
   		\includegraphics[width=2cm]{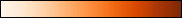}
    };
	\node[text width=3cm] at (9.95,-0.12) {$0$ cm};
	\node[text width=3cm] at (12.85,-0.12) {$2.5$ cm};	
	
	\node at (0.4,3.05) {(a)};
	\node at (3.45,3.05) {(b)};
	\node at (6.5,3.05) {(c)};
	\node at (9.55,3.05) {(d)};
	
%
\end{tikzpicture}
\caption{Liver displacement estimation. a) We present a discretized liver geometry to our convolutional neural network along with a zero-displacement boundary condition (red) and a visible surface area displacement (orange). b) The network estimates the displacement of every organ point (blue) from the known surface displacement. c) The actual displacement is known from a previously run simulation (blue, maximum magnitude of 7.3 cm). d) The error between the estimated displacement and the real displacement (visualized here for an internal slice through the liver) shows that points close to the visible surface area are displaced correctly while the maximum error of roughly 2.5 cm occurs on the opposite side of the organ.}
\label{Fig:LiverResultLargeDeformation}
\end{figure}
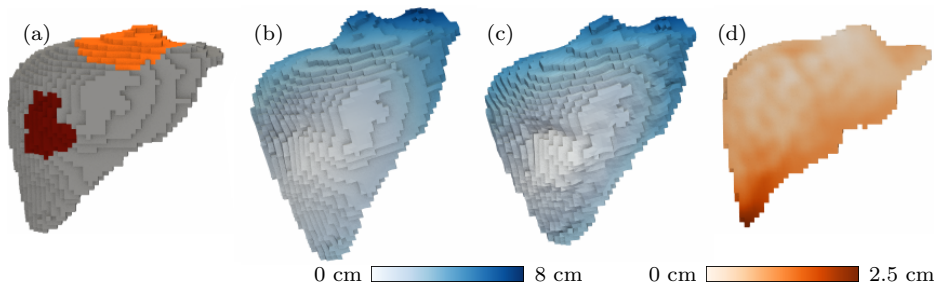

\section{Introduction}

An important prerequisite for soft tissue navigation during surgery is the simulation and prediction of tissue behavior. Here, the goal is to aid surgeons by visualizing information which is typically hidden - such as the position of tumors and blood-vessels - in a context-sensitive manner.

The usual workflow is to first generate a pre-operative model from CT or MRI data, containing structures of interest such as vessels and tumors. During the surgery, these structures are visualized for navigation. However, as organs deform during a surgery, the models have to be adapted to correctly reflect the intra-operative scene. This requires intra-operative sensor data such as video, CT or ultrasound to capture the state of the organs which can then be used to deform the pre-operative model so that it matches the current situation.
Since this task involves many sub-stages which should ideally all run in real-time (image processing, registration of organs, soft tissue simulation, rendering of virtual reality), it is desirable to use systems which require as little computational resources as possible.

When acquiring intra-operative sensor data to feed into the model, there are many challenges. Besides the limited field of view and noisy data, there are often unknown parameters which are either difficult or impossible to obtain during a surgery, such as knowledge about unseen organ surfaces, friction between organs and material parameters such as the tissue's elasticity. Approximations have to be used and the deformation model needs to be able to deal with uncertain information.

The real-time simulation of deformable elastic bodies is a complex problem and an active area of research.
Many pre- to intra-operative registration systems are based on the Finite Element Method (FEM) which can accurately calculate deformations of organs \cite{Suwelack2014,Reichard2017,Plantefeve2016,Mendizabal2017,PeterlikFastElRegistration,Simpson2012}. The FEM has several benefits like the possibility to precisely represent complex meshes by using small irregular volume elements and the ability to compute physically accurate solutions. It can be made real-time capable \cite{Wu2014,Allard11,Bui2018}, but due to the large, often sparse matrices the speed gain obtained by running the code on the GPU is limited. Changes in the mesh topology (such as a cut) require a computationally expensive update of the topology and the matrices \cite{Wu2014}. It is also difficult to incorporate unknown material parameters or uncertain boundary conditions \cite{Peterlik2017}.


Recently, convolutional neural networks (CNNs) have proven to be a powerful tool in building many data-driven models \cite{DeepLearning2016}, with the benefit of parallelizing well on modern GPUs.
In this work, we explore the usage of a CNN to estimate an organ's internal deformation from known surface deformation. The network is trained on synthetic FEM simulation data and learns to interpret the mesh structure of an organ as well as boundary conditions to calculate the displacement of internal points.  Our goal is to determine the displacement of internal structures when knowing about the displacement of (some) surface points. We assume that surface correspondences between pre-operative and intra-operative models have been computed using data from an intra-operative modality such as the laparoscope.


Our contribution consists of a novel method to generate a real-time capable soft tissue model which immediately generalizes to new patients, by training on a large number of synthetic meshes.
We show that our model performs well on both synthetic and real data even without knowing the precise nature of the underlying material model.
Since the training data makes very few assumptions about the structure of the organ and the type of deformation, our method shows that neural networks can indeed be taught how soft tissue deformation works in a general setting.
In effect, we substitute a highly engineered computational model for a data-driven one and in doing so we reach a very low computation time. Our code is available online\footnote{Code publically available at: \url{https://gitlab.com/nct_tso_public/cnn-deformation-estimation}}.

\subsection{Related Work}

The idea of using neural networks to estimate the outcome of an FEM simulation is not new. Hambli et al. \cite{Hambli2006} have used a fully connected net to predict the velocity and angle of a tennis ball after hitting a racket.
A similar approach has been adopted by Tonutti et al. \cite{Tonutti17} for predicting the movement of a tumor during brain surgery, but the displacement of the healthy tissue is not considered. Rechowicz et al. \cite{Rechowicz2013} estimate the displacement of a patient's rib cage, but only predict the displacement of surface nodes.
In contrast, Marooka et al. \cite{Morooka2008} estimate the displacement of a full liver model using neural networks by superimposing basic deformation modes. 
These approaches show that neural networks can indeed be trained to estimate soft tissue behavior. However, they work with the data of a single patient, requiring re-training (and in some cases even re-designing) the network for every new patient. Additionally, they use the acting surface forces as input which are very difficult to obtain intra-operatively.

Yamamoto et al. \cite{Yamamoto2017} show that neural networks can estimate the deformation of a liver from the known displacement of a partial surface. They report very small errors while using only 3\% of the liver surface, but also design their network for specific patients and evaluate their method on the same liver mesh that was used during training.

Approaches which generalize to new patients were made by Lorente et al. \cite{Lorente2017}, who estimate the liver's deformation from breathing motion using various machine learning methods and Martínez-Martínez et al. \cite{Martinez2017} who estimate breast compression. However, both methods focus on single scenarios where the main direction of the forces stay similar throughout all experiments.


Our method is inspired by the work of Guo et al. \cite{Guo2016}, who have used neural networks to estimate the results of fluid dynamics simulations for real-time applications. One major difference is that fluid dynamics are often computed on regular grids with a fixed number of cells which makes them easier to handle with neural networks. In contrast to this, we need to re-sample the irregular simulation domain before we pass the data to the network.


\section{Methods}

The goal of this work is to estimate the current displacement of an organ and its internal structures when given a) the pre-operative geometric model of the organ and b) the displacement of the visible part of the organ's surface as seen by an intra-operative sensor.
Since CNNs work best with a regular grid, we discretize all data at regularly spaced points. In doing so, our network's input becomes a cube of voxels and the network's output becomes a displacement field with a three-dimensional displacement vector for each of these voxels.

We sample the cubical volume of side length $L$ into a grid $G$ of $N \times N \times N$ points of interest. For each point $p \in G$ we determine three properties:
\begin{itemize}
\item Organ structure: To represent the mesh, we determine the \emph{signed distance} $s(p) \in \mathbb R$ for each point $p$ to the nearest surface of the organ in meters.
\item Visible displacement: We assume that the displacement of part of the organ surface is known, for example by tracking features with a (stereo-)laparoscope. For these surface points, we assign a \emph{visible displacement} vector $u_{vis} \in \mathbb R^3$. For all other points, $u_{vis}(p)=(0,0,0)^T$.
\item Zero displacement: We assign a binary value $z(p)$ which is set to one if a point of the surface is fixed to surrounding tissue and should be considered static. For all other points, $z$ is set to zero.
\end{itemize}

Knowing $s$, $u_{vis}$ and $z$, the goal is to find a displacement vector $u \in \mathbb R^3$ for each point $p$. The following sections explain how we randomly generate training data, how we build the network and how the network is trained to generate the displacement field $u$ for an organ.

\subsection{Synthetic Training Data} \label{Sec:TrainingData}

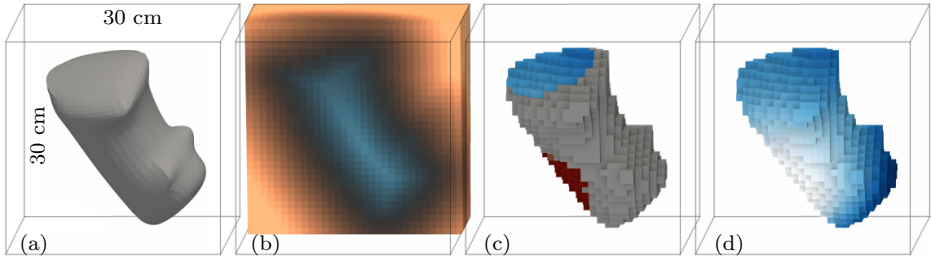
\begin{figure}
\begin{tikzpicture}patient-specific
	\node[anchor=south west,inner sep=0] at (0,0.4) {
   		\includegraphics[width=\textwidth]{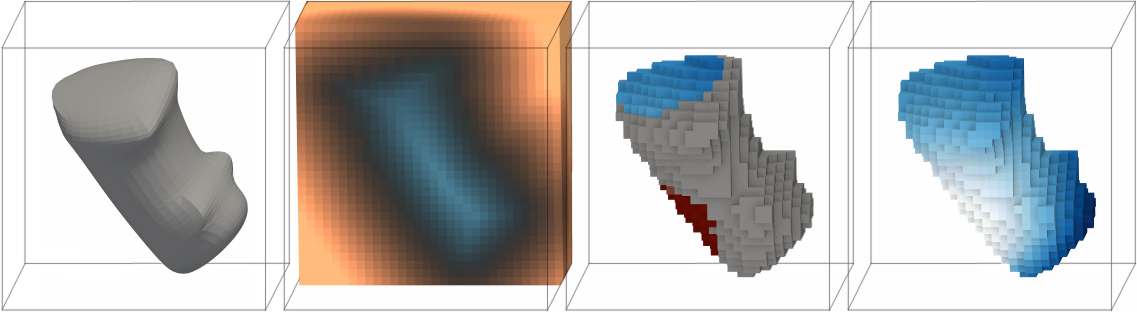}
    };
	
%
	
	\node at (0.4,0.55) {(a)};
	\node at (3.45,0.55) {(b)};
	\node at (6.5,0.55) {(c)};
	\node at (9.55,0.55) {(d)};
	
	\node at (1.7,3.57) {$30$ cm};
	\node[rotate=90] at (0.45,2) {$30$ cm};
\end{tikzpicture}
\caption{a) Random mesh structure, b) signed distance function $s$ on the grid $G$, c) visible displacement $u_{vis}$ (blue, magnitude) and zero displacement $z$ (red), d) target displacement $u_{tar}$ (magnitude). On b), half of the grid is hidden. In c) and d) (signed distance function $s > 0$) are clipped away to show only the organ structure.}
\label{Fig:Voxelization}
\end{figure}

Since training neural networks can be very time consuming, it is often not feasible to train a network before surgery using patient-specific data. Instead, we generate synthetic datasets which are based on simulations of random, organ-like meshes (see Fig. \ref{Fig:Voxelization}). First, a random surface mesh is generated by extruding and deforming a mesh primitive multiple times. The volume inside this surface is then filled with tetrahedral elements with the Gmsh \cite{GMSH} software.
We choose boundary conditions which are inspired by laparascopic surgeries:
A \emph{zero displacement} boundary condition is applied to a random surface region with a radius ranging from 2.5 cm to 5.5 cm (indicating areas where the organ is fixed to other organs) and a random force between 0 and 1 N is applied to another random surface region with a radius ranging from 1.5 to 2.5 cm (simulating instruments which manipulate the organ).

A non-linear, homogeneous, isotropic material model is chosen for the mesh, because we expect large deformations and expect to know very little about a patient's specific tissue parameters. The Youngs Modulus is set to 1.7 kPa and the Poisson Ratio to 0.35. The \emph{Elmer} simulation software \cite{Elmer2013} is used to run the steady-state simulations, resulting in a known target displacement vector $u_{tar}$ for each vertex (internal and external) of the random meshes.

The grid $G$ with side length $L = 30$ cm and $N = 64$ is generated and $z$ and $u_{tar}$ are calculated for each point $p$ via interpolation with a gaussian kernel.
To generate the visible displacement $u_{vis}$, another random surface region is selected and the target displacement vector $u_{tar}$ is copied to $u_{vis}$ for each point in the area (for other points, $u_{vis}(p)=(0,0,0)^T$). 

This process is repeated for 10 000 random meshes. If a randomly extruded mesh lies partly outside the 30 cm large cube or if it has self-intersections, the sample is discarded. The same is done if the deformation is larger than 10 cm, since these samples are much rarer and result in a very unevenly distributed dataset.
We augment the training data by flipping the grid along the X, Y and Z axis or any combination of the three, resulting in a factor eight increase in the number of training samples.
Of the resulting dataset, roughly 90\% (37440 samples) are used for training and the remaining data is used for validation.
Before passing the data to the network, we scale $s$ and $z$ by a factor of $0.1$ which improves convergence properties.

\subsection{Network Architecture}

The input to our CNN is the regular grid $G$. Each point $p \in G$ has five values assigned to it ($s$, $z$ and the three values from $u_{vis}$), so the input to the network contains $N \times N \times N \times 5$ scalars. Since we want to calculate the three components of $u$ for each $p$, the output of the network is $N \times N \times N \times 3$ scalars.

When learning displacement fields, a force acting on one side of the mesh usually influences not only the displacement in nearby points but can potentially have an effect on all organ points, even those on the far side of the organ. This means that in our network, each output must have the potential to be influenced by each input point (i.e. each output point must have a receptive field spanning the entire input).

The network uses an architecture similar to U-Net \cite{UNet2015}, with an encoder which reduces the input resolution and learns a high-level representation of the data and a decoder which reconstructs the high resolution output. We also use skip connections which copy features forward without modifying them, to allow the decoder to incorporate more detail in the computed output. Unlike U-Net, our network works with three-dimensional input data and all convolutions are calculated across the three dimensions.

The decoder uses average pooling layers to decrease the resolution of the input data, leading to a bottleneck with a resolution of $8^3$. In the bottleneck, the convolutional kernels are large enough to carry information across the computational domain, meeting the requirement for the large field of view in the output layers. The decoder has three upsampling layers, each of which performs a simple nearest-neighbour interpolation to double the resolution, ensuring that the network's output resolution is the same as the input resolution.
All convolutions have a kernel side length $k_s$ of 3 and padding of 1 and each is followed by a SoftSign non-linear activation function.

\begin{figure}
\includegraphics[width=\textwidth]{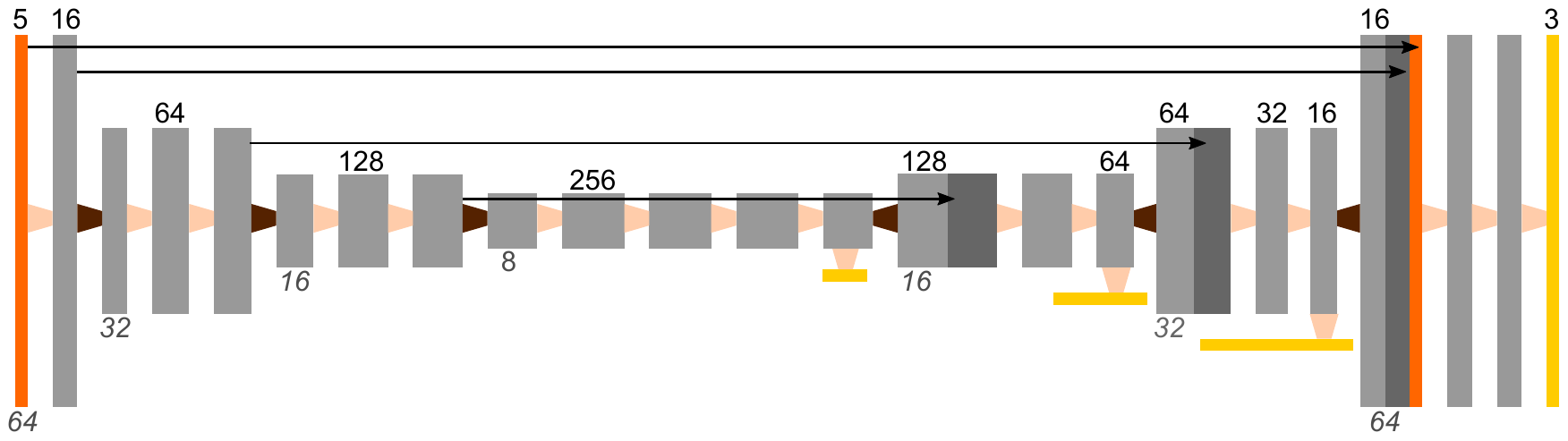}
\caption{Network architecture. Each bar represents the output or input to a convolutional layer. The network downsamples the input data to lower resolutions while increasing the number of channels per grid point. Then it increases resolution while decreasing the number of channels. Number of channels are shown above the bars and side lengths of the voxel grids are shown underneath (only changes are indicated). Skip-connections allow the network to access earlier information by copying feature maps to later layers. Some layers generate additional (downsampled) outputs (yellow bars).}
\label{Fig:NetworkArchitecture}
\end{figure}

For all points $p$ which lie outside the organ ($s(p) > 0$), the output $u_{est}(p)$ is set to zero. The final network architecture is depicted in Fig. \ref{Fig:NetworkArchitecture}. It has about 9.1 million learnable parameters.

\subsection{Error Functions}

Instead of only comparing the network's output to the target displacement, we let the network compute additional down-sampled displacement estimations and compare them to down-sampled version of the target displacement. This encourages the network to estimate correct displacements in the encoding- and bottleneck-layers while letting the decoding layers focus on increasing resolution.
Thus the network computes additional lower-resolution outputs $u_i$ at intermediate steps (compare Fig. \ref{Fig:NetworkArchitecture}) and a down-scaled version of the target displacement $u_{tar,i}$ is created for each resolution.
The mean square error is then calculated for each resolution $i$:

\begin{equation}
\mathcal{L}_i( u_i, u_{tar,i} ) = \frac{1}{N_i^3}\sum_{p}O(p) \, \lVert u_i(p) - u_{tar,i}(p)\rVert^2
\end{equation}
where $N_i^3$ is the number of points at resolution $i$, $O(p)$ is zero if p is outside the organ and one otherwise and $\|\centerdot\|$ denotes the magnitude of a vector. In practice,  $i \in \{64,32,16,8\}$.
The final error is the weighted sum of the errors over the four different resolutions:
\begin{equation}
\mathcal{L}( u, u_{tar} ) = \sum_{i} \lambda_{i} \, \mathcal{L}_i( u_i, u_{tar,i} )
\end{equation}
where the $\lambda$ are weighting factors. We choose $\lambda_{64}=\lambda_{32}=\lambda_{16}=\lambda_{8}=1$. The network is trained using the Adam \cite{Diederik15} optimizer until the error on the validation dataset no longer decreases.

\section{Experiments and Results}

We perform multiple experiments using the displacement estimation network.
In a first in-silico test, we test whether the network can generalize from the synthetic training structures to the shape of a real liver as segmented from CT data and how the network deals with varying amounts of visible surface.
Secondly, the network is tested on CT data of a phantom liver model undergoing a large deformation.
In a final test, we use human in-vivo data showing liver deformation due to breathing and let the network register the liver in the inhaled state to the liver in exhaled state.

All of these experiments are very different from the training data, as they contain never before seen mesh structures, material parameters and deformations. Thus, we implicitly evaluate the network's ability to generalize from the synthetic training data to settings which are closer to real-world scenarios.

Training as well as experiments were carried out on a personal computer with four Intel\textregistered \, i7 cores (4.20GHz) and an Nvidia GeForce GTX 1080 GPU with 8 GB of video RAM.

\subsection{In-Silico: Liver Mesh from CT}

We generate a new dataset with the same methods previously described for the training data (Section \ref{Sec:TrainingData}), but instead of a random mesh, we use the mesh of a patient's liver (OpenHELP Phantom \cite{Kenngott2015}). Again, random zero displacement and forces act on the organ and the Elmer software calculates displacements. After separating out samples with a displacement greater than 10 cm, this process results in 1334 deformed liver samples. Each of these samples has the same mesh structure but a different $u_{tar}$, $z$ and $u_{vis}$. The network runs on each sample, generating an estimated displacement field $u_{est}$ (see Fig. \ref{Fig:LiverResultLargeDeformation}).

\begin{figure}
\includegraphics[width=\textwidth]{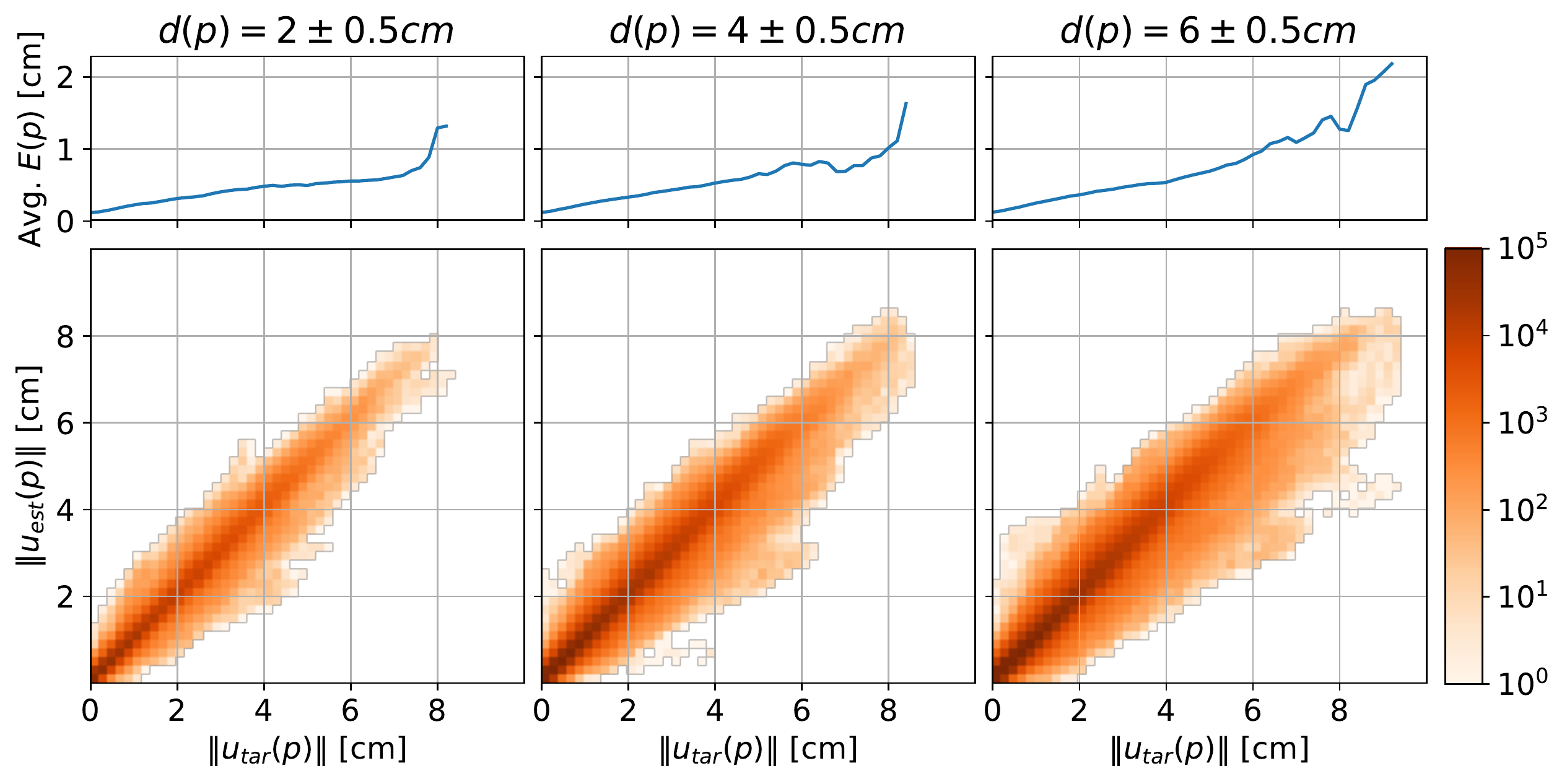}
\caption{Visualization of metrics for the deformed liver dataset, at three different depths (left column: $d(p) \approx 2 cm$, center column $d(p) \approx 4 cm$, right column: $d(p) \approx 6 cm$). For all six plots, the abscissa indicates the magnitude of $u_{tar}$, i.e. by how much a point \emph{should} be displaced. The top row of plots show how the average error increases with increasing target displacement as well as with increasing depth.
To generate the bottom row of plots, the points were sorted into bins of size 0.2 cm according to the magnitudes of their $u_{tar}$ and $u_{est}$ and the color of the plots indicates how many points end up in each bin. Most points (note the logarithmic scale) lie close to the diagonal, indicating that the network displaced them by the correct amount. As the distance from the visible area increases (rightmost plot), so does the average error. It can also be seen that the network is more likely to over-estimate the displacement than to under-estimate it. For all plots, points outside the organ ($s(p) > 0$) are ignored.}
\label{Fig:LiverErrorsByDisplacement}
\end{figure}

For each point $p$ inside the liver, a displacement error $E(p)$ can be calculated given the actual displacement $u_{tar}(p)$ and the estimated displacement $u_{est}(p)$ as 
$E(p) = \|u_{tar}(p)-u_{est}(p)\|$.
On average, this error is expected to increase as the distance to the visible displacement increases.
To quantify and visualize this effect, the points are sorted by their distance to the nearest visible surface point in the current sample. We call this distance the \emph{depth} $d(p)$ and collect all points from all 1334 data samples which have a similar depth. Fig. \ref{Fig:LiverErrorsByDisplacement} shows how the average errors increase with the depth as well the target displacement $u_{tar}(p)$.

\begin{figure}
\includegraphics[width=0.5\textwidth]{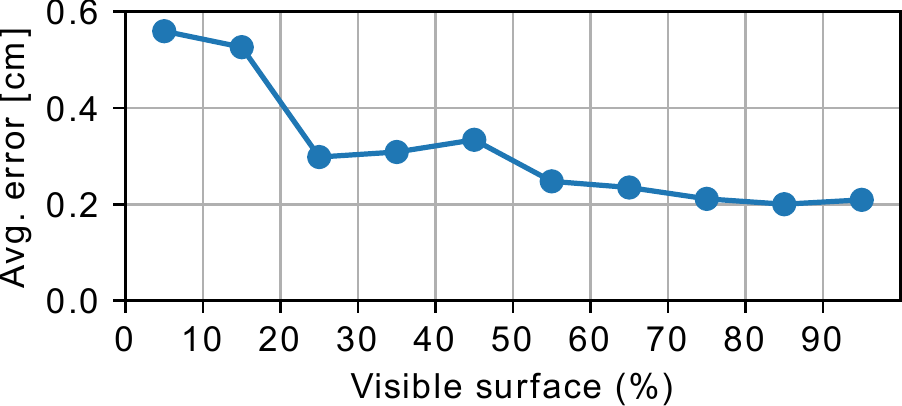}
\caption{Average error in the 1334 in silico liver samples by their amount of visible surface. To compute the averages, samples were first sorted into bins with a width of 10\% each. The largest drop in error is when approximately 20\% of the whole organ surface is visible. Incidentally, this corresponds to the amount of surface used in \cite{Suwelack2014} and is roughly the amount of surface which is accessible in a laparoscopic setting.}
\label{Fig:LiverErrorsByVisibleSurfaceArea}
\end{figure}

The average error is also expected to depend on the amount of visible surface.
To show how the error behaves as a larger percentage of surface becomes visible, we also sort the samples by percentage of visible surface and plot the average error (see Fig. \ref{Fig:LiverErrorsByVisibleSurfaceArea}).

\subsection{Phantom: Silicone Liver Deformation}
\begin{figure}
\begin{tikzpicture}
	\node[anchor=south west,inner sep=0] at (0,00) {
   		\includegraphics[width=\textwidth]{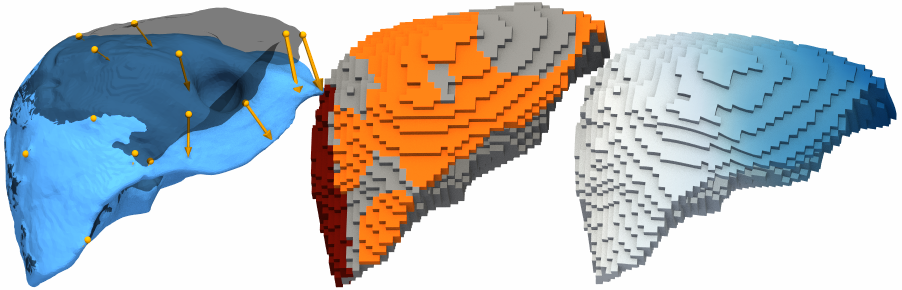}
    };
    \draw[->,ultra thick,white] (2.9,3.6) -- (3.0,2.8);
\end{tikzpicture}
\caption{Left: Original, undeformed CT model (grey) deformed by a strong force (white arrow), resulting in a large deformation (blue). Yellow arrows indicate the manually annotated sparse surface point correspondences.
Center: The surface areas where a visible displacement (orange, values interpolated from sparse surface dislacement) and zero displacement condition (red) are set.
Right: The network's estimated displacement field, applied to the points.}
\label{Fig:PhantomLiverResults}
\end{figure}

To test the network's ability to transfer its learned displacement estimation, we used the data of a silicone liver undergoing a large deformation (Liver registration dataset, Suwelack et al. \cite{Suwelack2014}, \emph{open-cas.org}).
The dataset contains the surface $S_O$ of the original, undeformed liver and a second surface $S_D$ which shows the same liver after it has been deformed by applying a strong force to its side using a spherical object (Fig. \ref{Fig:PhantomLiverResults}, left).
Six small Teflon markers were placed into the liver and their positions before and after the deformation can be used to determine a target registration error.
Due to the deformation, they move by up to 46.6 mm (mean 23.9 mm).

Again, we calculate the signed distance function $s$ for every grid point using the original surface $S_O$. The zero displacement condition $z$ is set for approximately 21\% of the liver surface where the organ was fixed to neighboring structures. To generate the visible displacement $u_{vis}$, we manually annotate 13 points on the anterior side of the surface $S_O$ and their correspondences on $S_D$. This sparse displacement information is then interpolated to other surface points, giving us an approximated dense surface displacement for roughly 18.5\% of the organ surface (Fig. \ref{Fig:PhantomLiverResults}, center).

We use the network to estimate the internal displacement of the organ.
The final registration error for the Teflon markers is 5.1 mm on average, with a maximum of 7.6 mm.
Suwelack et al. \cite{Suwelack2014} used 19\% of the surface area and also reported a mean error of 5.1 mm (maximum: 6.2 mm). While our method only takes a small fraction of the computation time (20 milliseconds compared to multiple seconds), we note that their method also solves the surface correspondence problem, which we assume as given.

\subsection{In-Vivo: Human Breathing Motion}

The network is applied to in-vivo human breathing motion data (3D-IRCADb-02 dataset, IRCAD, Strasbourg, \emph{ircad.fr/research/3d-ircadb-02/}). This dataset contains surface meshes of abdominal organs in an inhaled and exhaled state. Our goal is to register the liver in inhaled state to the exhaled (target) state.

First, we perform a rigid registration using the ICP algorithm \cite{ICP1992}.
To estimate the surface displacement, we then use the coherent point drift (CPD) algorithm \cite{CPD2006} on the liver surface model of the inhaled state to register it onto the liver model in exhaled state. 
We generate our grid $G$ from the inhaled state and set the zero displacement condition in the area where the vena cava touches the liver. The CPD's output is interpolated into G to generate the visible displacement $u_{vis}$.

\begin{figure}
\begin{tikzpicture}
	\node[anchor=south west,inner sep=0] at (0,00) {
   		\includegraphics[width=\textwidth]{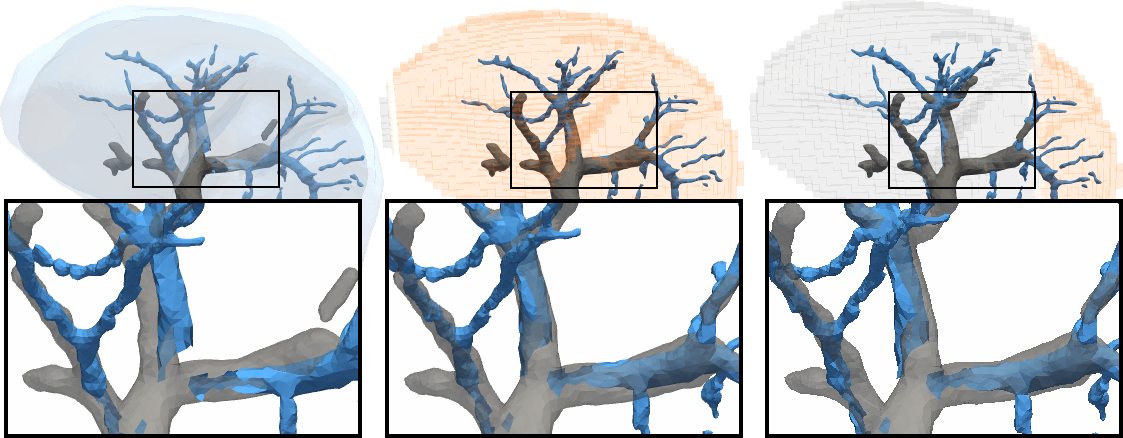}
    };
\end{tikzpicture}
\caption{Human breathing motion experiment. In all images, blue represents the source (inhaled) state and grey is the target (exhaled) state.
Left: After rigidly aligning the livers, there is still a non-rigid deformation of up to 2 cm, as seen in the liver surfaces (background) and the portal veins (zoomed box). Center: Using the full visible displacement from CPD, the network estimates a displacement field which is applied to the vessels, resulting in a good registration for all three main branches. Right: We decrease the amount of visible surface area (orange, shown here for 18\% of visible surface). The registration is still accurate close to the visible surface (right vessel branch) and becomes slightly less accurate as we go deeper into the liver.}
\label{Fig:HumanLiver}
\end{figure}

We iteratively decrease the amount of visible surface by moving a plane from the patient's posterior to their anterior and discarding all visible displacement information on the posterior side of the plane. We run the network on each of these steps, giving us a displacement field estimation for each amount of visible surface.
The generated displacement fields are used to deform the segmented portal vein tree inside the inhaled liver and compared to the portal veins of the exhaled liver. Since contrast agent was injected before the experiment, the two vessel trees differ a lot visually. This makes them difficult to compare quantitatively, but the results can be compared qualitatively (Fig. \ref{Fig:HumanLiver}).

\section{Discussion}

Our experiments show that the network can easily generalize to organ structures which it has never seen during training. In fact, when we tried further training the network with the patient specific mesh, no significant improvement was found. Even though we used the same material parameters for all training samples, the network performed well on evaluation data with different material properties.
The accuracy of the model depends mainly on the amount of visible surface and how close this visible area is to the area of largest deformation. This is in accordance with other research which indicates that the boundary conditions and a good geometric model have a higher influence on correct tissue modeling than the used material model \cite{Misra2009}\cite{Suwelack2011}.

The surface information available in the phantom experiment was very sparse, yet the very simple interpolation scheme we used to generate the dense surface displacement did not stop the network from creating satisfactory results. Similarly, in all experiments, the network deals well with the boundary conditions even though they are interpolated into our relatively coarse grid, indicating that it is not very susceptible to noise.

When training on purely artificial data, care must be taken that the data is representative of the real-world problem which the network should solve. We make multiple assumptions and simplifications which need to be addressed in the future:

During our discretization, we place a grid point roughly every 4.7 mm. While this resolution is similar to many real-time biomechanical models inside the organ, it results in a rough approximation of the organ surface.

Secondly, the training dataset does not simulate gravity and varying atmospheric pressure. The network shows promising results on the phantom and the human liver, both of which feature gravity, but adding body forces to the training dataset could further improve the network's ability to simulate real situations.

Thirdly, the network assumes that the zero displacement boundary condition is known. This is usually not the case in a real intra-operative setting, where only rough assumptions can be made. Our experience shows, however, that the model is robust to changes in size of the zero displacement area.

We also assume that surface correspondences between the pre- and intra-operative meshes are known. In the example of laparoscopic liver surgery, finding correspondences is far from trivial due to smooth, textureless surfaces and the large deformations. 
One approach would be to compute an initial non-rigid registration using a method such as \cite{Suwelack2014}, which matches the two surfaces without needing to compute features. Subsequent real-time tracking of intra-operative surface features \cite{Giannarou2013} could be used to keep the correspondences up to date.
To circumvent the difficulty of finding low-level geometric and texture features, current research focuses on using high-level cues - such as anatomical landmarks, the silhouette of the organ in the laparoscopic camera image and shading - to update a biomechanical model \cite{Adagolodjo2017,Koo2017}. Since these approaches compute displacement vectors to update their biomechanical model iteratively until they find a good registration, a method such as ours could be incorporated directly into a similar registration approach.



\section{Conclusion}

In this work, we employ a novel, data-driven model to estimate displacement information which is usually calculated using highly specialized hand-engineered models.
Due to the high speed, our method opens up many new possibilities in real-time applications. Besides being used in laparoscopic navigation, it could easily be extended to tackle brain-shift in neurosurgery or could potentially be used in motion-compensation during radio-therapy.
The method could also be used to estimate unknown boundary conditions and material parameters.

We have shown that a data-driven model can be used to model soft tissue deformation without seeing patient-specific data during the training phase. The model's accuracy is similar to other models, it is robust to simplifications in its input and it is very fast.

\section{Compliance with Ethical Standard}
\noindent \textbf{Conflict of Interest:} The authors declare that they have no conflict of interest.
\noindent This article does not contain any studies with human participants or animals performed by any of the authors.

\bibliographystyle{spmpsci}      
\bibliography{Literature}   

\end{document}